\documentclass[twoside,twocolumn,9pt]{article}
\usepackage{extsizes}
\usepackage[super,sort&compress,comma]{natbib} 
\usepackage[version=3]{mhchem}
\usepackage[left=1.5cm, right=1.5cm, top=1.785cm, bottom=2.0cm]{geometry}
\usepackage{balance}
\usepackage{times,mathptmx}
\usepackage{sectsty}
\usepackage{graphicx} 
\usepackage{lastpage}
\usepackage[format=plain,justification=justified,singlelinecheck=false,font={stretch=1.125,small,sf},labelfont=bf,labelsep=space]{caption}
\usepackage{float}
\usepackage{fancyhdr}
\usepackage{fnpos}
\usepackage[english]{babel}
\addto{\captionsenglish}{%
  \renewcommand{\refname}{Notes and references}
}
\usepackage{array}
\usepackage{droidsans}
\usepackage{charter}
\usepackage[T1]{fontenc}
\usepackage[usenames,dvipsnames]{xcolor}
\usepackage{setspace}
\usepackage[compact]{titlesec}
\usepackage{hyperref}

\usepackage{epstopdf}

\definecolor{cream}{RGB}{222,217,201}

\begin{document}

\pagestyle{fancy}
\thispagestyle{plain}
\fancypagestyle{plain}{

\fancyhead[C]{\includegraphics[width=18.5cm]{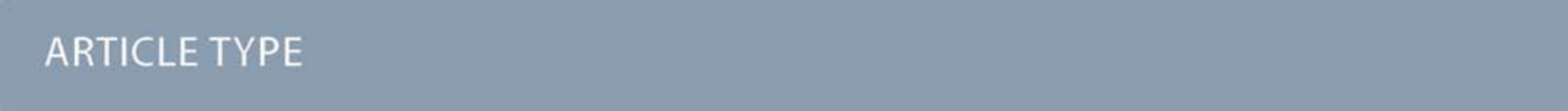}}
\fancyhead[L]{\hspace{0cm}\vspace{1.5cm}\includegraphics[height=30pt]{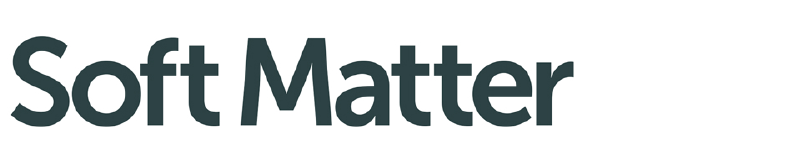}}
\fancyhead[R]{\hspace{0cm}\vspace{1.7cm}\includegraphics[height=55pt]{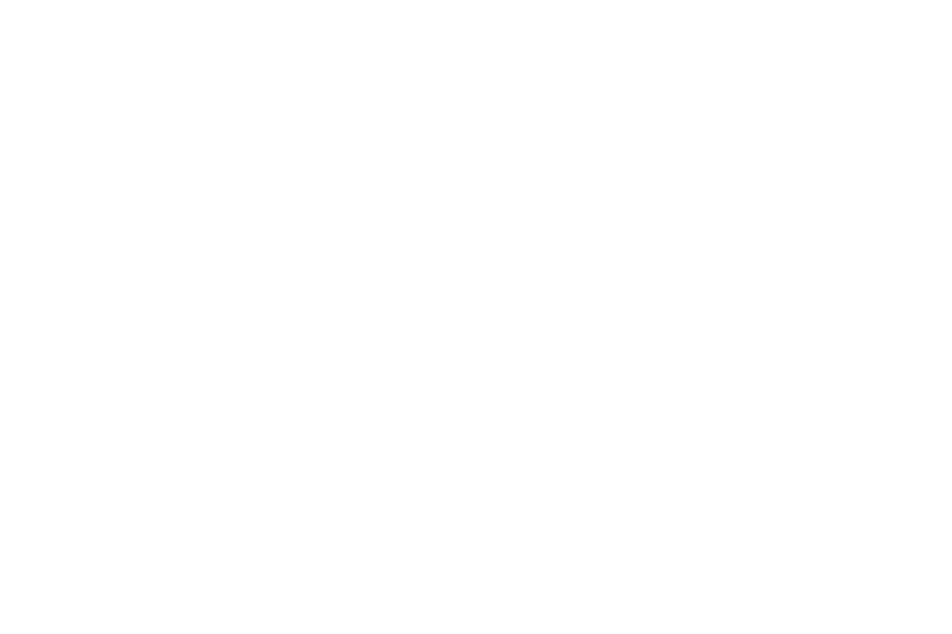}}
\renewcommand{\headrulewidth}{0pt}
}

\makeFNbottom
\makeatletter
\renewcommand\LARGE{\@setfontsize\LARGE{15pt}{17}}
\renewcommand\Large{\@setfontsize\Large{12pt}{14}}
\renewcommand\large{\@setfontsize\large{10pt}{12}}
\renewcommand\footnotesize{\@setfontsize\footnotesize{7pt}{10}}
\makeatother

\renewcommand{\thefootnote}{\fnsymbol{footnote}}
\renewcommand\footnoterule{\vspace*{1pt}%
\color{cream}\hrule width 3.5in height 0.4pt \color{black}\vspace*{5pt}} 
\setcounter{secnumdepth}{5}

\makeatletter 
\renewcommand\@biblabel[1]{#1}            
\renewcommand\@makefntext[1]%
{\noindent\makebox[0pt][r]{\@thefnmark\,}#1}
\makeatother 
\renewcommand{\figurename}{\small{Fig.}~}
\sectionfont{\sffamily\Large}
\subsectionfont{\normalsize}
\subsubsectionfont{\bf}
\setstretch{1.125} 
\setlength{\skip\footins}{0.8cm}
\setlength{\footnotesep}{0.25cm}
\setlength{\jot}{10pt}
\titlespacing*{\section}{0pt}{4pt}{4pt}
\titlespacing*{\subsection}{0pt}{15pt}{1pt}

\fancyfoot{}
\fancyfoot[LO,RE]{\vspace{-7.1pt}\includegraphics[height=9pt]{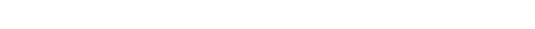}}
\fancyfoot[CO]{\vspace{-7.1pt}\hspace{13.2cm}\includegraphics{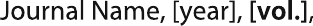}}
\fancyfoot[CE]{\vspace{-7.2pt}\hspace{-14.2cm}\includegraphics{RF}}
\fancyfoot[RO]{\footnotesize{\sffamily{1--\pageref{LastPage} ~\textbar  \hspace{2pt}\thepage}}}
\fancyfoot[LE]{\footnotesize{\sffamily{\thepage~\textbar\hspace{3.45cm} 1--\pageref{LastPage}}}}
\fancyhead{}
\renewcommand{\headrulewidth}{0pt} 
\renewcommand{\footrulewidth}{0pt}
\setlength{\arrayrulewidth}{1pt}
\setlength{\columnsep}{6.5mm}
\setlength\bibsep{1pt}

\makeatletter 
\newlength{\figrulesep} 
\setlength{\figrulesep}{0.5\textfloatsep} 

\newcommand{\topfigrule}{\vspace*{-1pt}%
\noindent{\color{cream}\rule[-\figrulesep]{\columnwidth}{1.5pt}} }

\newcommand{\botfigrule}{\vspace*{-2pt}%
\noindent{\color{cream}\rule[\figrulesep]{\columnwidth}{1.5pt}} }

\newcommand{\dblfigrule}{\vspace*{-1pt}%
\noindent{\color{cream}\rule[-\figrulesep]{\textwidth}{1.5pt}} }

\makeatother

\twocolumn[
  \begin{@twocolumnfalse}
\vspace{3cm}
\sffamily
\begin{tabular}{m{4.5cm} p{13.5cm} }

\includegraphics{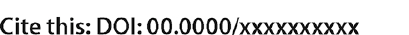} & \noindent\LARGE{\textbf{Jammed packings of 3D superellipsoids with tunable packing fraction, contact number, and ordering$^\dag$}} \\
\vspace{0.3cm} & \vspace{0.3cm} \\

 & \noindent\large{Ye Yuan,\textit{$^{a,b}$} Kyle VanderWerf,\textit{$^{c}$} Mark D. Shattuck,\textit{$^{d}$} and Corey S. O'Hern$^{\ast}$\textit{$^{b,c,e,f}$}} \\

\includegraphics{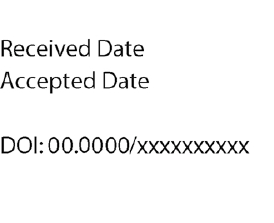} & \noindent\normalsize{We carry out 
numerical studies of static packings of frictionless superellipsoidal 
particles in three spatial dimensions. We consider more than $200$ different 
particle shapes by varying the three shape 
parameters that define superellipsoids. We characterize the structural and 
mechanical properties of both disordered and ordered packings using two 
packing-generation protocols. We perform athermal quasi-static compression 
simulations starting from either random, dilute configurations (Protocol 1) 
or thermalized, dense configurations (protocol $2$), which allows us 
to tune the orientational order of the packings. In general, we find 
that the contact numbers at jamming onset for superellipsoid packings are hypostatic, 
with $z_J < z_{\rm iso}$, where $z_{\rm iso} = 2d_f$ and $d_f = 5$ 
or $6$ depending on whether the particles are axi-symmetric or not. 
Over the full range of orientational order, we find that 
the number of quartic modes of the dynamical matrix for the packings always 
matches the number of missing contacts relative to the isostatic value. 
This result suggests that there are no 
mechanically redundant contacts for ordered, yet hypostatic packings 
of superellipsoidal particles. Additionally, we find that the packing 
fraction at jamming onset for diordered packings of superellipsoidal 
particles can 
be collapsed using two particle shape parameters, e.g. the asphericity 
$\mathcal{A}$ and 
reduced aspect ratio $\beta$ of the particles.} 

\\

\end{tabular}

 \end{@twocolumnfalse} \vspace{0.6cm}

  ]

\renewcommand*\rmdefault{bch}\normalfont\upshape
\rmfamily
\section*{}
\vspace{-1cm}


\footnotetext{\textit{$^{a}$~Department of Mechanics and Engineering Science, College of Engineering, Peking University, Beijing 100871, China; E-mail: yuanyepeking@pku.edu.cn}}
\footnotetext{\textit{$^{b}$~Department of Mechanical Engineering and Materials Science, Yale University, New Haven, Connecticut 06520, USA. }}
\footnotetext{\textit{$^{c}$~Department of Physics, Yale University, New Haven, Connecticut 06520, USA. }}
\footnotetext{\textit{$^{d}$~Benjamin Levich Institute and Physics Department, The City College of New York, New York, New York 10031, USA. }}
\footnotetext{\textit{$^{e}$~Department of Applied Physics, Yale University, New Haven, Connecticut 06520, USA. }}
\footnotetext{\textit{$^{f}$~Graduate Program in Computational Biology and Bioinformatics, Yale University, New Haven, Connecticut 06520, USA. }}




\section{Introduction}

Athermal particulate materials, such as granular media, foams, and
emulsion droplets, typically jam, or become solid-like with a non-zero
static shear modulus when they are compressed to sufficiently large
packing fractions.~\cite{o2002random, van2009jamming, parisi2010mean, liu2010jamming}  Unwanted jamming occurs in many industrial
processes, such as clogging in hopper flows,~\cite{thomas2015fraction} and controlled jamming
and unjamming has been used in robotics to grip soft, sharp, or fragile
objects.~\cite{jaeger2015celebrating} Further, unjamming in geological systems, such as landslides 
and earthquakes, causes significant financial and human loss.  

Many prior studies have focused on jamming in model systems composed
of frictionless, spherical particles.  Disordered packings of
frictionless, monodisperse spherical particles are isostatic at
jamming onset with $z_J = z_{\rm iso}$ contacts per particle, where
$z_{\rm iso} = 2 d_{f}=6$ and $d_{f} = 3$ is the number of translational
degrees of freedom for spheres, and with packing fraction at jamming onset
$\phi_J \approx 0.64$.~\cite{makse2000packing, o2002random, van2009jamming} Previous work has characterized the critical
scaling of the structural and mechanical properties~\cite{o2002random, ellenbroek2006critical, olsson2007critical, wang2013critical} and the anomalous
vibrational density of states~\cite{silbert2005vibrations, wyart2005effects} of jammed packings of spherical
particles.~\cite{liu2010jamming}

However, most athermal, particulate systems in industrial processes
and in nature are composed of highly non-spherical
particles.~\cite{torquato2010jammed, lu2015discrete} In general,
disordered jammed packings of non-spherical particles are hypostatic
with $z_J < z_{\rm iso}$, where $z_{\rm iso}=10$ or $12$ for
axisymmetric and non-axisymmetric particles,
respectively.~\cite{donev2004improving, donev2007underconstrained,
  wouterse2009contact, jiao2010distinctive} Thus, disordered jammed
packings can possess a range of contact numbers, $6 \le z_J \le 12$,
and packing fractions at jamming onset that depend on the shape of the
constituent particles. In two spatial dimensions (2D), we showed
recently that disordered packings generated via athermal, quasistatic
compression for a wide variety of non-spherical shapes are
mechanically stable, despite the fact that $z_J < z_{\rm
  iso}$.~\cite{vanderwerf2018hypostatic} We found that certain types
of contacts between nonsperical particles can constrain mulitple
degrees of freedom and that the number of missing contacts below the
isostatic value matches the number of quartic eigenmodes of the
dynamical matrix.  At jamming onset, perturbing the system along a
quartic eigenmode causes the total potential energy to increase as the
fourth power (not quadratically) in the perturbation
amplitude.~\cite{mailman2009jamming, schreck2012constraints}

Given that jammed packings of non-spherical particles can occur over a
wide range of contact numbers and packing fractions, is it possible to
{\it a priori} determine whether a system is jammed if we are only
given its $z$ and $\phi$?  For {\it disordered} packings of
monodisperse spheres, we know that if $\phi > 0.64$ and $z > 6$, the
packing is jammed. For disordered packings of convex-shaped particles
in 2D, we found that the packing fraction at jamming onset can be
collapsed approximately onto a master curve that depends only on the
shape parameter $A = p^2/{4\pi a}$, where $p$ is the perimeter and $a$
is the area of the particle.~\cite{vanderwerf2018hypostatic} In 2D, $\phi_J
\approx 0.84$ for $A=1$, $\phi_J$ increases with $A$, reaching a peak
near $A \approx 1.1$, and then decreases continuously with further
increases in $A$. Results for $\phi_J$ have also been reported for packings 
of nonspherical particles in 3D, but separately for each family of shapes, 
e.g., ellipsoids,~\cite{donev2007underconstrained}
spherocylinders,~\cite{wouterse2007effect, wouterse2009contact, zhao2012dense} and
spheropolyhedra.~\cite{yuan2019random} Here, we will address the 
question of whether there is a  general relationship
between the packing fraction at jamming onset and one or more particle
shape parameters in packings of non-spherical particles in 3D.

Further, few studies have attempted to connect the contact number to
mechanical stability for {\it ordered} packings of non-spherical
particles,~\cite{borzsonyi2013granular, asencio2017experimental} despite the fact that packings of monodisperse particles
that deviate by less than $20\%$ from perfect sphericity can possess
significant translational and orientational order. In particular, does
the relationship between the number of missing contacts below the
isostatic value and number of quartic modes hold for ordered or
partially ordered packings of non-spherical particles?  One might
expect that some of the ``extra'' contacts that occur in ordered
packings, may be mechanically redundant,~\cite{lopez2013jamming} and
therefore will not contribute to the packing's stability, resulting in
a mismatch between the number of missing contacts and the number of
quartic modes.

\begin{figure}
\centering
  \includegraphics[height=5cm]{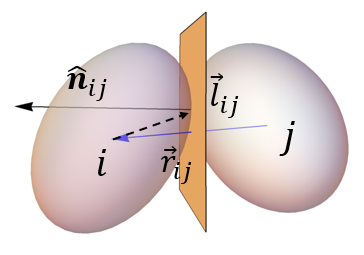}
\caption{Illustration of the simulation model for two contacting 
superellipsoids $i$ and $j$. ${\hat n}_{ij}$ is the unit normal to the tangent 
plane at the point of contact (pointing toward particle $i$), ${\vec r}_{ij}$
is the center-to-center vector between particles $i$ and $j$, and 
${\vec l}_{ij}$ is the vector from the center of particle $i$ to the point 
of contact between particles $i$ and $j$.}
\label{contact}
\end{figure}

We investigate these questions by generating static packings of monodisperse
frictionless, superellipsoidal-shaped particles in 3D using numerical
simulations.~\cite{delaney2010packing, zhao2017particle,
  yuan2018coupling, liu2018uniform} We consider more than $200$
different particle shapes by changing the shape parameters that define
superellipsoids. For each packing, we determine $\phi_J$, $z_J$, the
orientational order, and the eigenvalues and eigenmodes of the
dynamical matrix. We carry out two packing-generation protocols. In
Protocol $1$, we jam the packing via athermal quasistatic
compression,\cite{o2002random, schreck2012constraints,
  smith2010athermal, smith2014variable} starting from a random, dilute initial
configuration of particles. In Protocol $2$, we thermalize an unjammed
configuration at an intermediate packing fraction before applying the
same athermal quasistatic compression protocol (Protocol $1$). We find
that Protocol $1$ generates globally disordered packings with a narrow
distribution of jammed packing fractions and contact numbers. Protocol
2, on the other hand, is able to generate packings of superellipsoidal
particles with a wide range of orientational order.

We find several key results. First, for disordered packings of
superellipsoidal particles in 3D generated via Protocol 1, we find
that the jammed packing fraction can be collapsed onto a master curve
using two shape parameters instead of only one as we found for
2D.~\cite{vanderwerf2018hypostatic}  In addition, we find that the
number of contacts, even in {\it ordered} packings of superellipsoids,
determines their mechanical stability. In particular, the number of
quartic eigenmodes of the dynamical matrix matches the number of
missing contacts relative to the isostatic value in ordered
superellipsoid packings, as well as in disordered packings.

\begin{figure}
\centering
  \includegraphics[height=9cm]{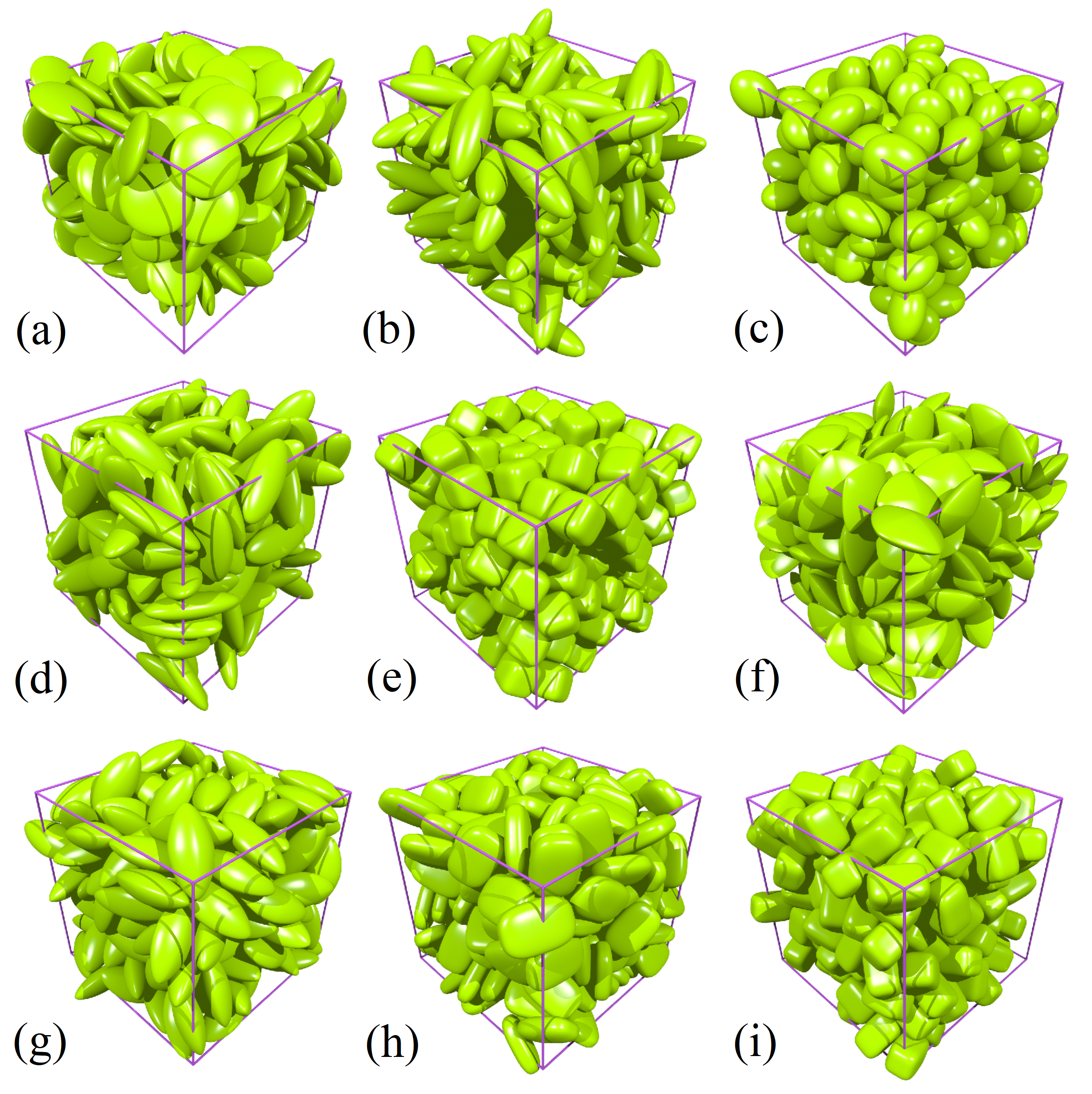}
\caption{Examples of nine static packings of superellipsoid particles with 
different shapes. The particle shape is characterized by $(p,w_1,w_2)$: 
(a) oblate ellipsoid $(1,0.3,1)$, (b) prolate ellipsoid $(1,1,3)$, (c) 
self-dual ellipsoid $(1,0.8,1.25)$, (d) general ellipsoid $(1,0.6,2.36)$, 
(e) superball $(2,1,1)$, and four superellipsoids with (f) $(0.75,0.4,1)$, 
(g) $(0.85,0.7,2)$, (h) $(1.5,0.5,1.5)$, and (i) $(2,1,1.5)$. }
\label{examples}
\end{figure}

The article is divided into several sections. In Sec.~\ref{methods},
we review the definition of superellipsoids, describe the two
packing-generation protocols we implement, and define the
orientational order parameters we use to measure the degree of order
in jammed packings. In Sec.~\ref{results}, we present our key
results. Finally, in Section 4, we summarize our results and discuss
directions for future research.  We also include three Appendices.  In
Appendix~A, we show that we widely sample the two shape
parameters that characterize the shape of superellipsoids. In
Appendix~B, we examine the {\it local} orientational order in 
superellipsoid packings. Finally, in Appendix~C, we show the
correlation between the average curvature of the particles at
interparticle contacts and the average contact number for packings of
superellipsoids.

\section{Methods}
\label{methods}

In this section, we begin by defining the shape parameters for
superellipsoids, and explain the wide variation in particle shape that
is possible by tuning these parameters. Next, we describe our two
protocols, the athermal protocol 1, and the thermal protocol 2, which
we use to generate disordered and ordered jammed packings of these
shapes, respectively. We then discuss calculations of the eigenvalues
and eigenmodes of the dynamical matrix for superellipsoid packings to
measure their mechanical response. Finally, we define the two order
parameters that we use to quantify the orientational order in the
packings.

\subsection{Model of superellipsoidal particles}

The surface of a superellipsodal particle located at the origin is defined by  
\begin{equation}
\label{eom}
  |x/a|^{2p} + |y/b|^{2p} + |z/c|^{2p} = 1,
\end{equation}
where $a$, $b$, and $c$ ($a \le b \le c$) are the lengths of the
semi-major axes, and $p$ is the deformation
parameter.~\cite{jiao2010distinctive,ni2012phase} For superellipsoids,
there are three independent parameters that control the particle
shape, i.e., $p$ and the two aspect ratios $w_1=a/b$ and $w_2=c/b$. If $a
= b$, there is only one relevant asepct ratio $w = c/a$ and if $b =
c$, $w = a/c$. Note that the particle shape reduces to a
superball when $a = b = c$. By tuning $p$, we can vary the
superellipsoid shape from ellipsoidal ($p=1$) to octahedral ($p < 1$)
and cuboidal ($p > 1$).  We focus our studies on five specific $p$-values: $p =
0.75$, $0.85$, $1$, $1.5$, and $2$.

\begin{figure}
\centering
\includegraphics[height=3.4cm]{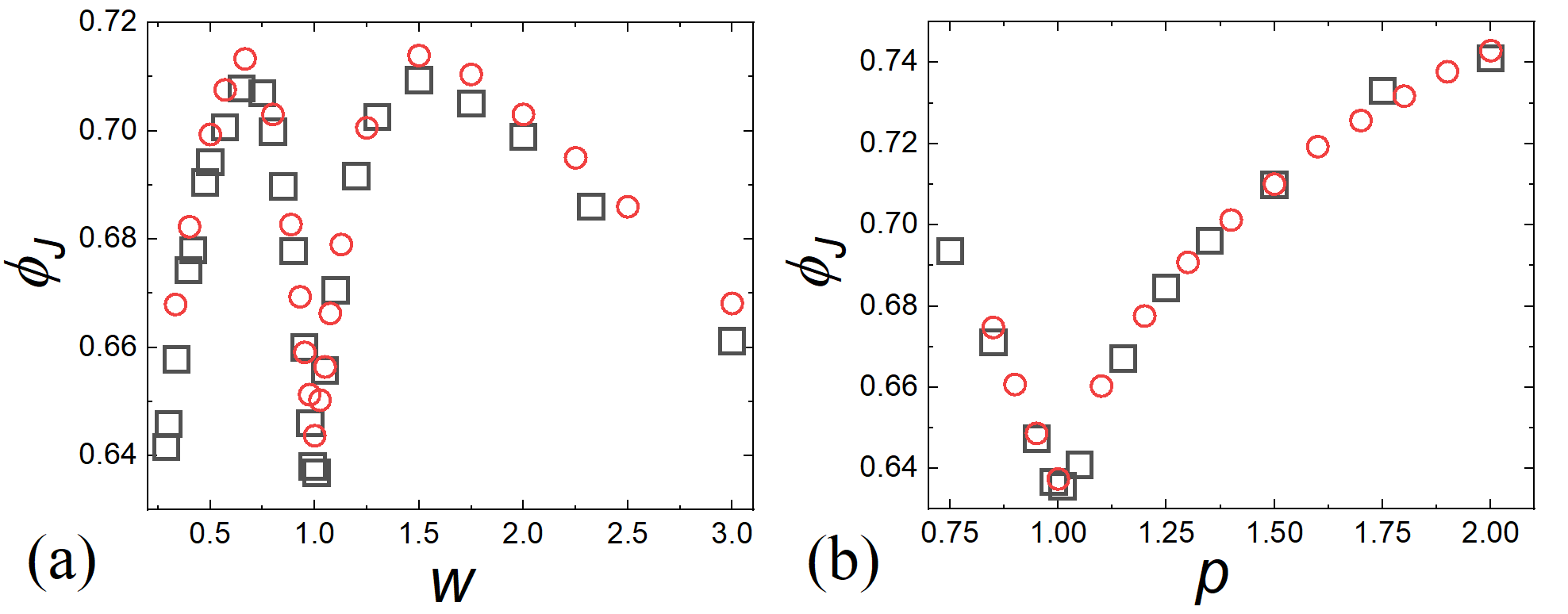}
\caption{(a) The packing fraction at jamming onset $\phi_J$ for 
packings of $N=400$ prolate ($a=b$) or oblate ($b=c$) 
spheroids generated using protocol $1$ versus the aspect 
ratio $w$ (open squares), as well $\phi_J$ for packings of spheroids 
from recent studies by Donev, {\it et al.}\cite{donev2007underconstrained} 
(open circles). (b) $\phi_J$ for packings of superballs ($a=b=c$) generated 
using protocol $1$ versus the deformation parameter $p$, as well as $\phi_J$ 
for packings of superballs from Jiao, {\it et al.}\cite{jiao2010distinctive}}
\label{prior}
\end{figure} 

Instead of $p$ and the aspect ratios, $w_1$ and $w_2$, the shape of
superellipsoids can also be characterized by $p$, the reduced aspect ratio
$\beta=ac/b^2$, and asphericity,
\begin{equation}
  \mathcal{A} = 1 - (4\pi)^{1/3}(3V_p)^{2/3}/A_p,
\end{equation}
where $V_p$ and $A_p$ give the particle volume and surface
area.\cite{zou1996evaluation, yu1996modifying} The shape parameter
$\beta$ allows us to distinguish ``flattened'' ($\beta < 1$) versus
``elongated'' ($\beta > 1$) shapes. The shape with $\beta = 1$ is termed
a self-dual ellipsoid, which shows anomalous properties in disordered\cite{donev2007underconstrained} and dense\cite{donev2004unusually}
packings. The asphericity satisfies $0 < {\cal A} < 1$, and ${\cal A}=1$ for
spheres. For the $p$ values studied, the superellipsoidal particle
shape in the $\beta$-${\cal A}$ plane is roughly bounded by the values
for prolate $\beta_{\rm max}({\cal A})$ and oblate ellipsoids
$\beta_{\rm min}({\cal A})$ as shown in Fig.~\ref{beta} in
Appendix~A.  We focus on ${\cal A}$-values from $0$ to $\sim
0.35$ and sample $\beta_{\rm min}({\cal A}) < \beta < \beta_{\rm
  max}({\cal A})$.

\begin{figure}
\centering
  \includegraphics[height=9cm]{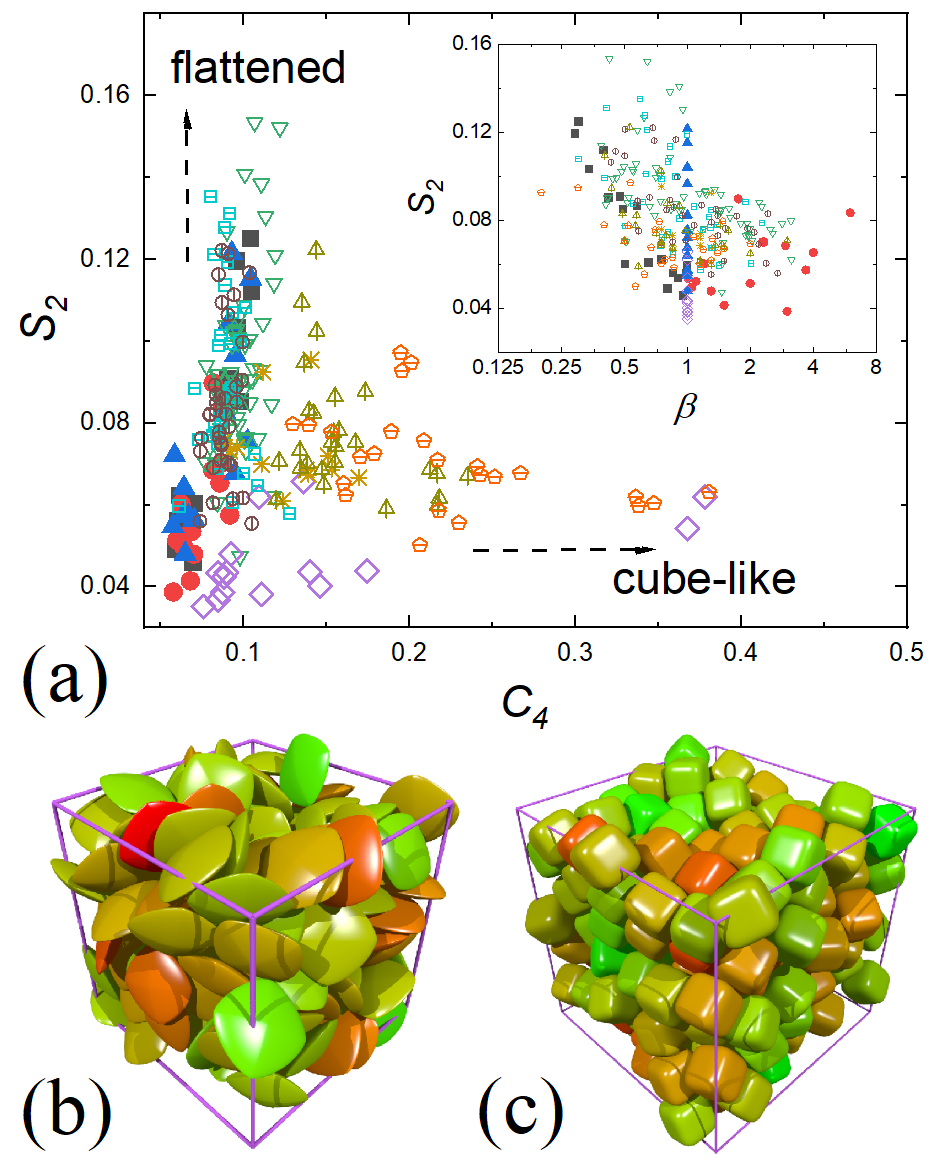}
\caption{(a) A scatter plot of the global nematic $S_2$ and cubatic $C_4$ 
order parameters for packings of superellipsoids generated via 
protocol $1$. The particle shapes include oblate ellipsoids 
(filled squares), prolate ellipsoids (filled circles), self-dual ellipsoids 
(filled upward triangles), general ellipsoids (downward open triangles), 
superballs (open diamonds), nearly spherical particles with $p\sim 1$ and 
$w\sim 1$ (asterisks), and $p=0.75$ (squares with lines), $0.85$ (circles with 
lines), $1.5$ (upward triangles with lines), and $2.0$ (pentagons with lines).
The vertical (horizontal) arrow indicates packings 
with increasingly flatter (cube-like) shapes. The inset shows a scatter 
plot of $S_2$ versus the normalized aspect ratio $\beta$ for the same data 
set. (b) Example
packing of superellipsoids with $p = 0.75$ and $w = 0.3$ and global nematic 
order $S_2 = 0.11$. (c) Example packing of superballs with $p = 2$ and 
global cubatic order $C_4 = 0.37$. We show the (b) local  
nematic and (c) local
cubatic order by coloring the particles with increasing local order 
from green to red.}
\label{scatter}
\end{figure} 

\begin{figure}
\centering
  \includegraphics[height=6cm]{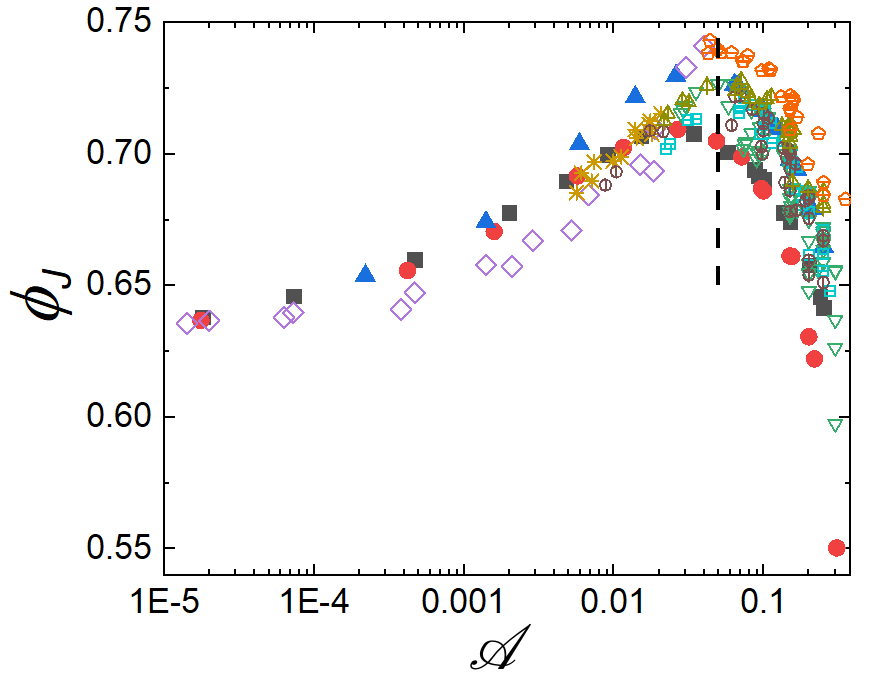}
\caption{The packing fraction at jamming onset $\phi_J$ versus the 
asphericity $\mathcal{A}$ for packings of the same shapes described in 
Fig.~\ref{scatter} generated via protocol $1$.  
The vertical dashed line marks the characteristic ${\cal A}_c \sim 0.05$ of 
the peak in the $\phi_J({\cal A})$.}
\label{phij}
\end{figure}

We consider pairwise, purely repulsive interactions between
superellipsoids using the Perram and Wertheim
formulation.\cite{perram1985statistical, donev2007underconstrained, schreck2012constraints,
  ni2012phase, marschall2019orientational}. For each pair of
overlapping superellipsoids $i$ and $j$, we calculate the volume
scaling factor $\eta_{ij}$ that brings the two superellipsoids to
exact tangency. The potential energy for particles $i$ and $j$ is then
defined by $U_{ij} = \epsilon \zeta_{ij}^2/2$, where $\epsilon$ is the
characteristic energy scale, $\zeta_{ij} = \eta_{ij}^2 - 1$, and
$\eta_{ij} \le 1$. The total potential energy is given by
$U=\sum_{i>j} U_{ij}$. The repulsive force on particle $i$ from $j$
${\vec f}_{ij} = {\vec \nabla}_i U$ is given by
\begin{equation}
  \vec{f}_{ij} = 2\epsilon \zeta_{ij}\eta_{ij}\hat{n}_{ij}/(\vec{r}_{ij}
\cdot\hat{n}_{ij})
\end{equation}
where $\hat{\emph{\textbf{n}}}_{ij}$ is unit normal of the tangent
plane between just-touching superellipsoids pointing toward $i$ and
${\vec r}_{ij}$ is the center-to-center vector pointing from
superellipsoid $j$ to $i$. The torque ${\vec \tau}_{ij}$ on particle $i$ from
$j$ is calculated using
\begin{equation}
\label{torque}
{\vec \tau}_{ij} = {\vec l}_{ij} \times {\vec f}_{ij},
\end{equation}
where ${\vec l}_{ij}$ is the vector from the center of particle $i$ to
the point of adjacency between superellipsoids $i$ and $j$.  See
Fig.~\ref{contact} for an illustration of ${\hat n}_{ij}$, ${\vec
  r}_{ij}$, and ${\vec l}_{ij}$ for two contacting superellipsoids. We
will measure lengths, energies, and forces in terms of $a$,
$\epsilon$, and $\epsilon/a$.

\begin{figure}[t!]
\centering
\botfigrule
  \includegraphics[height=10cm]{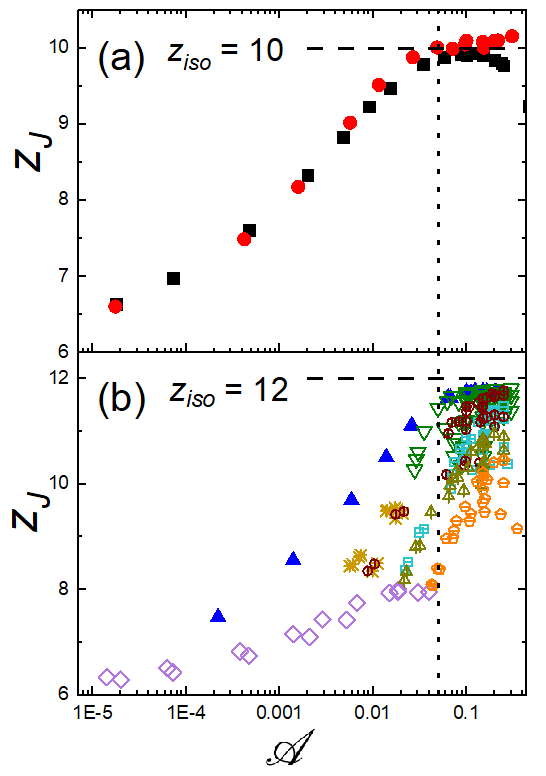}
  \caption{The contact number  at jamming onset $z_J$ versus asphericity 
${\cal A}$ for packings generated via protocol $1$ for (a) spheroids and 
(b) all other shapes. The symbols are the same as those used 
in Fig.~\ref{scatter}. The horizontal dashed lines in (a) and (b) 
indicate $z_{\rm iso} = 10$ and $12$ for the respective families of shapes. 
The vertical dotted line marks the threshold in 
$\mathcal{A}\sim$ 0.05 above which $z_J({\cal A})$ reaches a plateau for 
spheroids.}
\label{zij}
\end{figure}

\subsection{Packing-generation protocols}

We generate jammed packings of $N=400$ frictionless, monodisperse
superellipsoidal particles in cubic simulation cells with periodic
boundary conditions using two compression protocols: 1) an athermal
protocol and 2) a thermal protocol.  For protocol 1, we first
initialize an overlap-free, dilute configuration of particles with
random positions and orientations. We then compress the configuration
in small increments of packing fraction, $\Delta \phi=10^{-3}$,
minimizing the total potential energy $U$ using the L-BFGS
method\cite{liu1989limited} after each compression step. We terminate
the energy minimization procedure when the average normalized force on
a particle is below a small threshold, $\langle |\sum_j {\vec f}_{ij}|
\rangle/\langle f_{ij} \rangle < \Delta$, where $\Delta = 10^{-4}$.
We stop compressing the system when the total potential energy per
particle first satisfies $U/N > U_{\rm tol}$, where $U_{\rm tol} =
10^{-10}$.  We then measure the packing fraction $\phi_J$, contact
number $z_J$, and other quantities of the first jammed packing with
$U/N > U_{\rm tol}$ that is closest to $U_{\rm tol}$.  We find that
the results presented here do not depend on the thresholds $\Delta$
and $U_{\rm tol}$.  Examples of nine static packings of
superellipsoids with different shapes generated via protocol 1 are
shown in Fig.~\ref{examples}.

For protocol 2, we first thermalize unjammed configurations at
intermediate packing fractions $\phi_i \sim 0.55$, between the freezing
and melting packing fractions for hard superellipsoids,\cite{ni2012phase, odriozola2012revisiting} using Monte
Carlo methods that do not allow particle overlaps for $N_s$ steps. We then 
input these configurations into the compression and energy 
minimization procedure described in protocol $1$. By varying 
$N_s$ and $\phi_i$, we can obtain jammed packings of superellipsoids with 
tunable $\phi_J$, contact number $z_J$, and degree of orientational 
order. 

To calculate average quantities for the packing fraction, contact
number, and other quantities at jamming onset, we average over $5$ to
$10$ independent initial conditions. We validated our methods for
generating jammed packings of superellipsoids by comparing our results
for $\phi_J$ from protocol 1 to those from recent studies of packings
of spheroids and superballs by Donev, {\it et
  al.}\cite{donev2007underconstrained, jiao2010distinctive} (See
Fig.~\ref{prior}.)

\subsection{Dynamical matrix}

The dynamical matrix, which provides all possible second derivatives
of the total potential energy with respect to the rotational and
translational degrees of freedom of the system, determines the linear
mechanical response of jammed particle packings. We define the
dynamical matrix as
\begin{equation}
\label{dm}
M_{kl} = \partial^2U/\partial{\xi_k}\partial{\xi_l},  
\end{equation}
where ${\vec \xi} = \{x_1, y_1, z_1, a\theta_1, a\phi_1, a\psi_1,\ldots,
x_N, y_N, z_N, a\theta_N, a\phi_N, a\psi_N\}$, ($x_i$,$y_i$,$z_i$) is the
location of the center of particle $i$, and
($\theta_i$,$\phi_i$,$\psi_i$) are the rotation angles about the $x$-,
$y$-, and $z$-axes used to define the orientation of particle $i$.
Thus, the dimension of the dynamical matrix is $6N\times 6N$.  For
jammed superellipsoid packings (in cubic simulation cells with
periodic boundary conditions), the dynamical matrix possesses $6N'-3$
nonzero eigenvalues $\lambda_i$ (with corresponding unit eigenvectors ${\hat
  e}_i$), where $N'=N-N_r$ and $N_r$ is the number of rattler
particles with unconstrained translational or rotational degrees of
freedom.

To determine $M_{kl}$, we calculated the first-order derivatives of
the dynamical matrix, $\partial{U}/\partial{\xi_k}$ analytically, and
calculated all of the the second-order derivatives numerically. We
find that the eigenvalues of the dynamical matrix do not depend
sensitively on the numerical derivatives for displacements $<
10^{-8}$.

\subsection{Order parameters}

In packings of non-spherical particles, one can measure the degree of
order in the translational (i.e. positions of the particle centers)
and rotational (i.e. orientations of the particles) degrees of
freedom.  In the systems we study, when the particle orientations are
ordered, the particle positions also contain significant order.  Thus,
in these studies, we will focus on quantifying the orientational
order.
 
We measure the global nematic
$S_2$\cite{veerman1992phase,wouterse2007effect} and cubatic $C_4$ order
parameters.\cite{john2008phase} $S_2$ is defined as the largest
eigenvalue of the $3\times 3$ matrix:
\begin{equation}
\label{s2}
S_{\alpha \beta} = \frac{3}{2}\langle {\hat s}_{\alpha i} 
{\hat s}_{\beta j}\rangle - \frac{\delta_{\alpha \beta}}{2}
\end{equation}
where $\delta_{\alpha \beta}$ is the Kronecker delta, $\alpha$,
$\beta=x$, $y$, and $z$, ${\hat s}_{\alpha i}$ is the $\alpha$-compoent
of the unit vector that characterizes the orientation of particle $i$. 
and $\langle . \rangle$ indicates an average over all pairs of particles
$i$ and $j$. 
${\hat s}_i$ is chosen as the shortest (longest) axis of the particle
when $\beta < 1$ ($\beta > 1$).  With this definition of ${\hat s}_i$, 
$S_2$ can capture stacking order that can occur in packings of flat shapes, 
as well as nematic order that can occur in packings of elongated shapes. 
$S_2 = 0$ for systems without orientational order and $1$ for systems 
with complete particle alignment.  

The cubatic order parameter~\cite{john2008phase} $C_4$ is obtained 
by first calculating the fourth-order Legendre polynomial,  
\begin{equation}
P_4({\hat t},{\hat u}_i) =\frac{1}{8}\left( 35 [\hat{\emph{\textbf{t}}}\cdot{\hat{\emph{\textbf{u}}}}_i]^4 - 30[\hat{\emph{\textbf{t}}}\cdot{\hat{\emph{\textbf{u}}}}_i]^2 + 3 \right),
\end{equation}
where ${\hat t}$ is the unit vector aligned with one of the $3N$
orientations of the semi-major axes of each of the particles and
${\hat u}_i$ is a unit vector aligned with one of the three
orientations of the semi-major axes for particle $i$. For each
particle $i$ in a given jammed packing, we select the ${\hat u}_i$
that maximizes $P_4({\hat t}, {\hat u}_i)$ for a given ${\hat t}$. We
then average $P_4^{\rm max}({\hat t})$ over all particles for a given
${\hat t}$ and define $C_4$ as the maximum over all $3N$ orientations
${\hat t}$. For $C_4 \sim 1$, packings possess large cubatic order, which
can occur in packings of cube-like particles with $p>1$. In
Appendix~B, we show results for the local nematic and cubatic order
in packings of superellipsoids.

\begin{figure}[t!]
\centering
  \includegraphics[height=13cm]{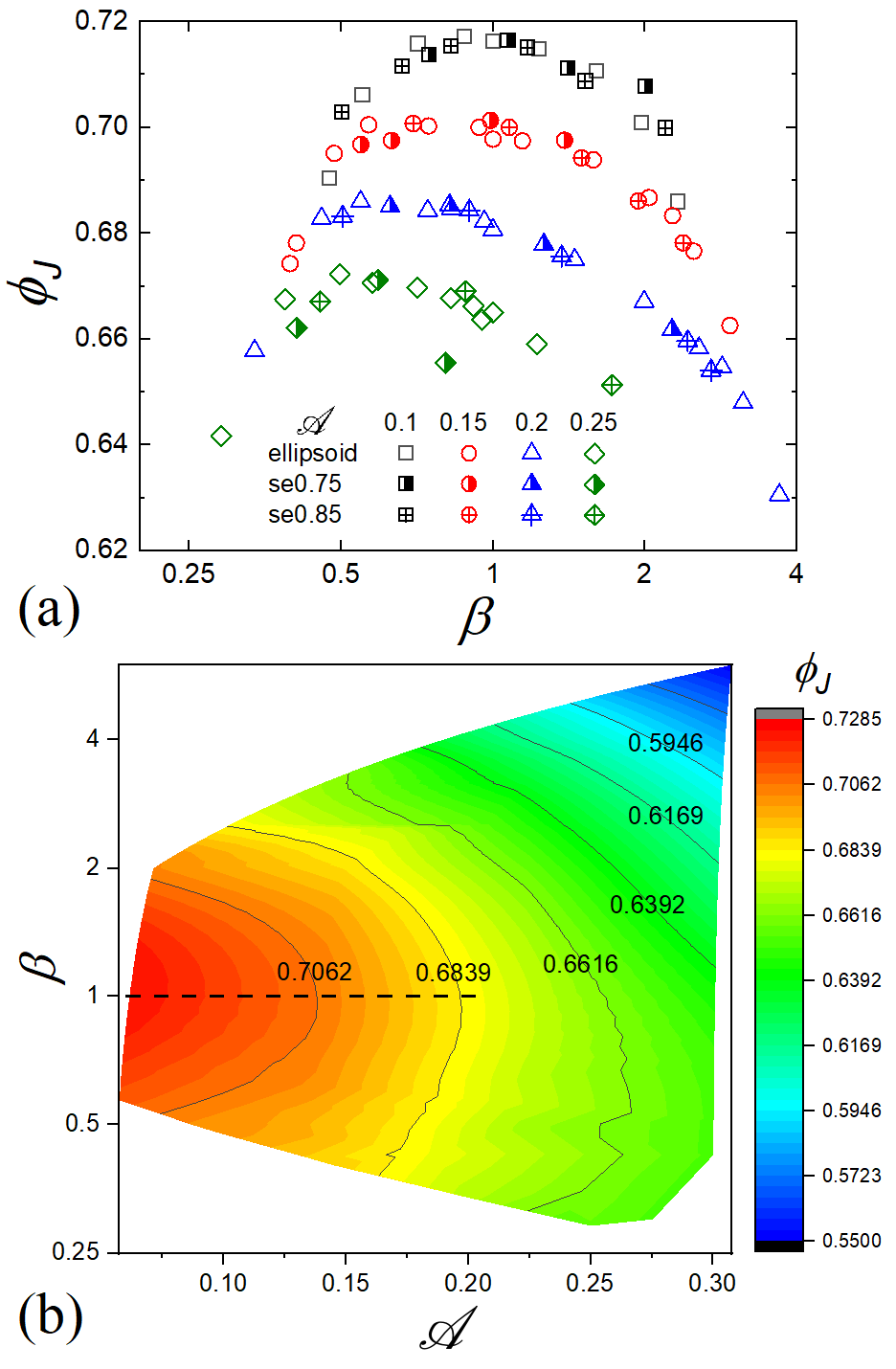}
\caption{(a) Packing fraction at jamming onset $\phi_J$ versus the 
reduced aspect ratio $\beta$ for packings of superellipsoids generated 
using protocol $1$. The plot includes ellipsoids with four values of the 
asphericity 
${\cal A} = 0.1$, $0.15$, $0.2$, and $0.25$ and two families of 
superellipsoids with $p = 0.75$ (se$0.75$) and 0.85 (se$0.85$). (b) Contour 
plot of $\phi_J$ as a function of ${\cal A}$ and $\beta$. The horizontal 
dashed line indicates $\beta = 1$.}
\label{contour}
\end{figure}

\section{Results and Discussion}
\label{results}

Our results are divided into two subsections. In
Sec.~\ref{disordered}, we present our results for disordered packings
of superellipsoids generated via protocol $1$. We show the global
nematic and cubatic order parameters for packings containing a wide
variety of superellipsoidal shapes. We find that the packing fraction
at jamming onset for disordered packings of superellipsoids can be
collapsed as a function of the two shape parameters, ${\cal A}$ and
$\beta$. In Sec.~\ref{ordered}, we show that we can tune the packing
fraction and contact number at jamming onset by increasing the
orientational order of the packings generated via protocol $2$. We
also show that, even for ordered packings, the number of quartic modes
of the dynamical matrix is equal to the isostatic number of contacts
minus the number of contacts in the packing.  Thus, we find
a direct link between the contact number and mechanical properties
even for ordered packings of superellipsoids.

\begin{figure*}
\centering
  \includegraphics[height=7.5cm]{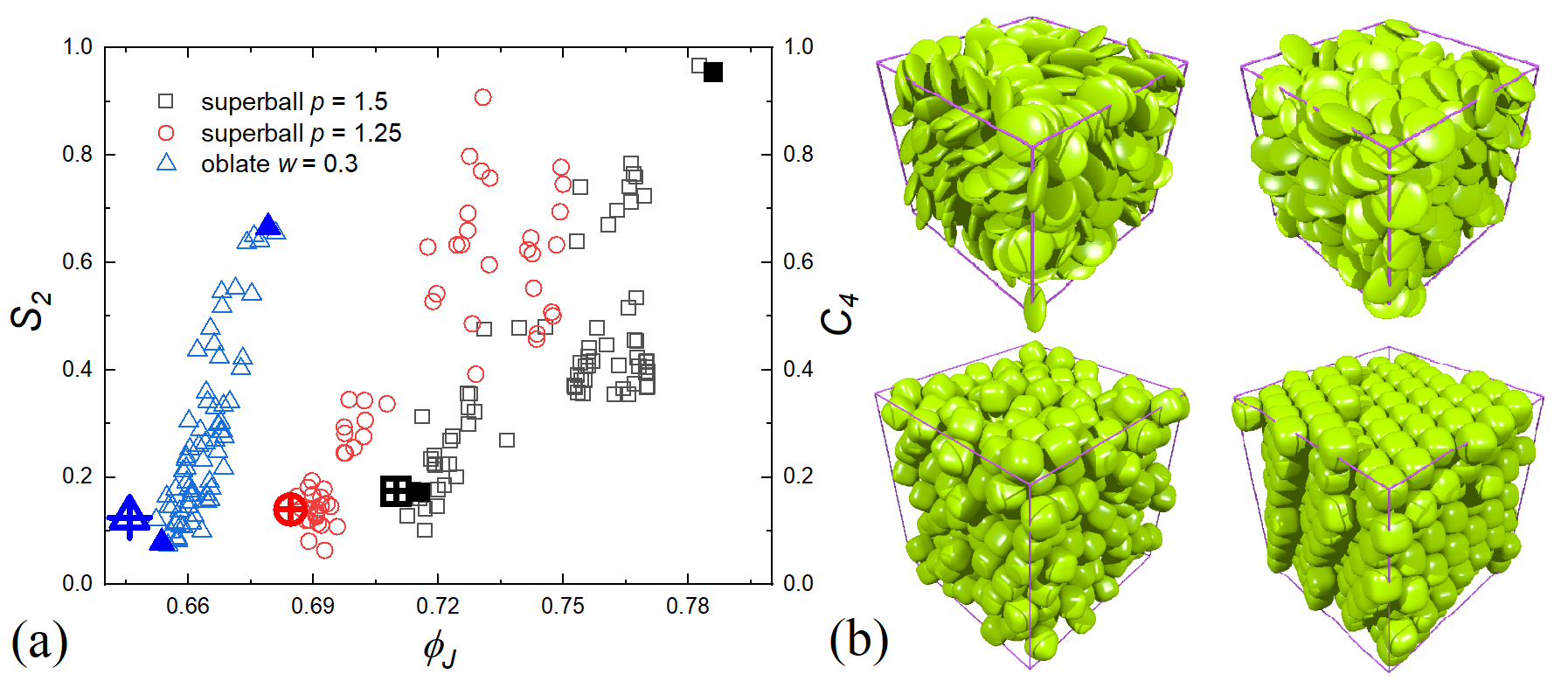}
\caption{(a) Gloabl nematic $S_2$ (left axis) and cubatic $C_4$ (right axis) 
order parameters for single packings of superballs
(with $p = 1.25$ and $1.5$) and oblate ellipsoids (with $w = 0.3$) 
generated using protocol $2$ are plotted versus $\phi_J$. The average values 
of $S_2$ and $C_4$  
for an ensemble of packings with the same particle shape, but generated 
using protocol $1$ 
are shown using corresponding symbols with crosses on the inside. The filled 
symbols represent the four packings shown in panel (b). (b) [top] Example packings 
generated via (left) protocol $1$ and (right) $2$ for 
oblate ellipsoids with $w=0.3$ and [bottom] example packings generated 
via (left) protocol $1$ and (right) $2$ for superballs with $p=1.5$.}
\label{protocol2}
\end{figure*}

\subsection{Disordered packings of superellipsoids}
\label{disordered}

In this section, we focus on the structural propreties of
superellipsoid packings generated via protocol $1$. In
Fig.~\ref{scatter} (a), we show a scatter plot of the global nematic
$S_2$ and cubatic $C_4$ order parameters for all jammed packings
generated using protocol $1$. We find that many of the packings are
disordered with $S_2$ and $C_4 \sim 1/\sqrt{N} \sim 0.07$.  However,
as demonstrated in the inset to Fig.~\ref{scatter} (a), $S_2$
increases as $\beta$ decreases below $1$ and the particle shape
flattens.  For more elongated shapes with $\beta >1$, $S_2$ is roughly
independent of $\beta$. We also find that the cubatic order increases
as the particles become more cube-shaped with $p>1$, even though the
packings were generated using the athermal protocol.  In
Fig.~\ref{scatter} (b) and (c), we show example packings of flattened
and cube-like superellipsoids generated via protocol $1$ wih elevated
values of $S_2$ and $C_4$. In (b), we show the local nematic order of
the particles for a packing of flattened superellipsoids with $p=0.75$
and $w=0.3$. In (c), we show the local cubatic order of the particles
for a packing of superballs with $p=2$. These packings possess local 
nematic and cubatic order. (See Appendix~B.)

In Fig.~\ref{phij}, we show the packing fraction at jamming onset
$\phi_J$ as a function of the asphericity ${\cal A}$ for a variety of
superellipsoid shapes.  The relation between $\phi_J$ and
${\cal A}$ is similar to that for packings of noncircular particles in
2D.\cite{vanderwerf2018hypostatic} $\phi_J$ starts at a relatively low
value for spherical particles (i.e. random close packing for
monodisperse spheres with $\phi_J(0) \approx 0.64$), $\phi_J$ grows with
increasing asphericity, reaching a peak $\phi_J \sim 0.70$-$0.74$ near
${\cal A} \sim 0.05$, and then $\phi_J$ begins decreasing, falling
below $\phi_J(0)$ for ${\cal A} > 0.1$. We also note that the data for
$\phi({\cal A})$ does not collapse as well onto a single curve in 3D,
compared to the collapse of $\phi_J({\cal A})$ for packings of 2D
noncircular particles.\cite{vanderwerf2018hypostatic}

In Fig.~\ref{zij}, we show the contact number at jamming onset $z_J$
versus the asphericity ${\cal A}$ for (a) spheroids with an axis of
symmetry and $z_{\rm iso} = 10$ and for (b) all other particle shapes
with $z_{\rm iso} = 12$. $z_J = 6$ for isostatic packings of spherical
particles in the limit ${\cal A} \rightarrow 0$. As found previously,
$z_J$ for packings of nonspherical particles does not jump
discontinuously from $6$ to $z_{\rm iso}$ when ${\cal A}$ increases
above zero. Instead, $z_J$ increases continuously with ${\cal
  A}$.  $z_J$ for some of the particle shapes reaches $z_{\rm iso}$ for ${\cal
  A} < 0.35$, {\it e.g.} oblate, prolate, self-dual, and general
elliposoids, but others, such as superellipsoids with $p=0.75$,
$0.85$, $1.5$, and $2.0$ do not. Note that $z_{\rm iso}$ is smaller
for spheroids, compared to $z_{\rm iso}$ for other non-axisymmetric
particle shapes, and thus the maximumum packing fraction for spheroids
is smaller than that for the other shapes we studied. We correlate 
values of $z_J < z_{\rm iso}$ for superellipsoids with the curvature 
at interparticle contacts in Appendix~C.

The packing fraction at jamming onset $\phi_J$ for packings of
superellipsoids does not completely collapse when plotted versus a
single shape parameter, {\it e.g.} the asphericity ${\cal A}$.  (See
Fig.~\ref{phij}.) This result suggests that $\phi_J$ for packings of
nonspherical particles in 3D depends on two or more shape
parameters. In Fig.~\ref{contour} (a), we show $\phi_J$ versus the
reduced aspect ratio $\beta$ for several values of the asphericity
${\cal A} =0.1$, $0.15$, $0.2$, and $0.25$, excluding cube-like
superellipsoids with $p>1$. All of the curves $\phi_J(\beta)$ are
concave down for the different values of ${\cal A}$. In
Fig.~\ref{contour} (b), we show a contour plot of $\phi_J$ as a
function of both $\beta$ and ${\cal A}$. We find that at small ${\cal
  A}$, the largest $\phi_J$, $\phi_J^{\rm max}$, occurs near $\beta =
1$, however, $\phi_J^{\rm max}$ shifts to $\beta >1$ when ${\cal A} >
0.2$. Thus, $\phi_J$ depends on both shape parameters ${\cal A}$ and
$\beta$.

\subsection{Tunable hypostaticity}
\label{ordered}

In this section, we show that we can increase the nematic or cubatic
order in packings of superellipsoids using protocol $2$ to generate
the packings. We compare the packing fraction and contact number at
jamming onset for packings generated via protocols $1$ and $2$. We
focus on packings of superballs with $p = 1.25$ and $1.5$ and
packings of oblate ellipsoids with $w = 0.3$.

In Fig.~\ref{protocol2} (a), we show the global nematic $S_2$ and
cubatic $C_4$ order parameters versus the packing fraction at jamming
onset $\phi_J$ for single packings of oblate ellipsoids (with $w=0.3$)
and superballs (with $p=1.25$ and $1.5$) generated via protocol
$2$. We also compare these results to those for packings of the same
shapes, but generated using protocol $1$. Example packings are displayed
in Fig.~\ref{protocol2} (b).  We find that $S_2$ and $C_4 < 0.1$-$0.2$
for packings generated via protocol $1$. However, $S_2$ and $C_4$ can
become larger than $0.7$ for packings generated using protocol
$2$. For all shapes studied, $\phi_J$ increases with increasing
orientational order.

\begin{figure}
\centering
  \includegraphics[height=11cm]{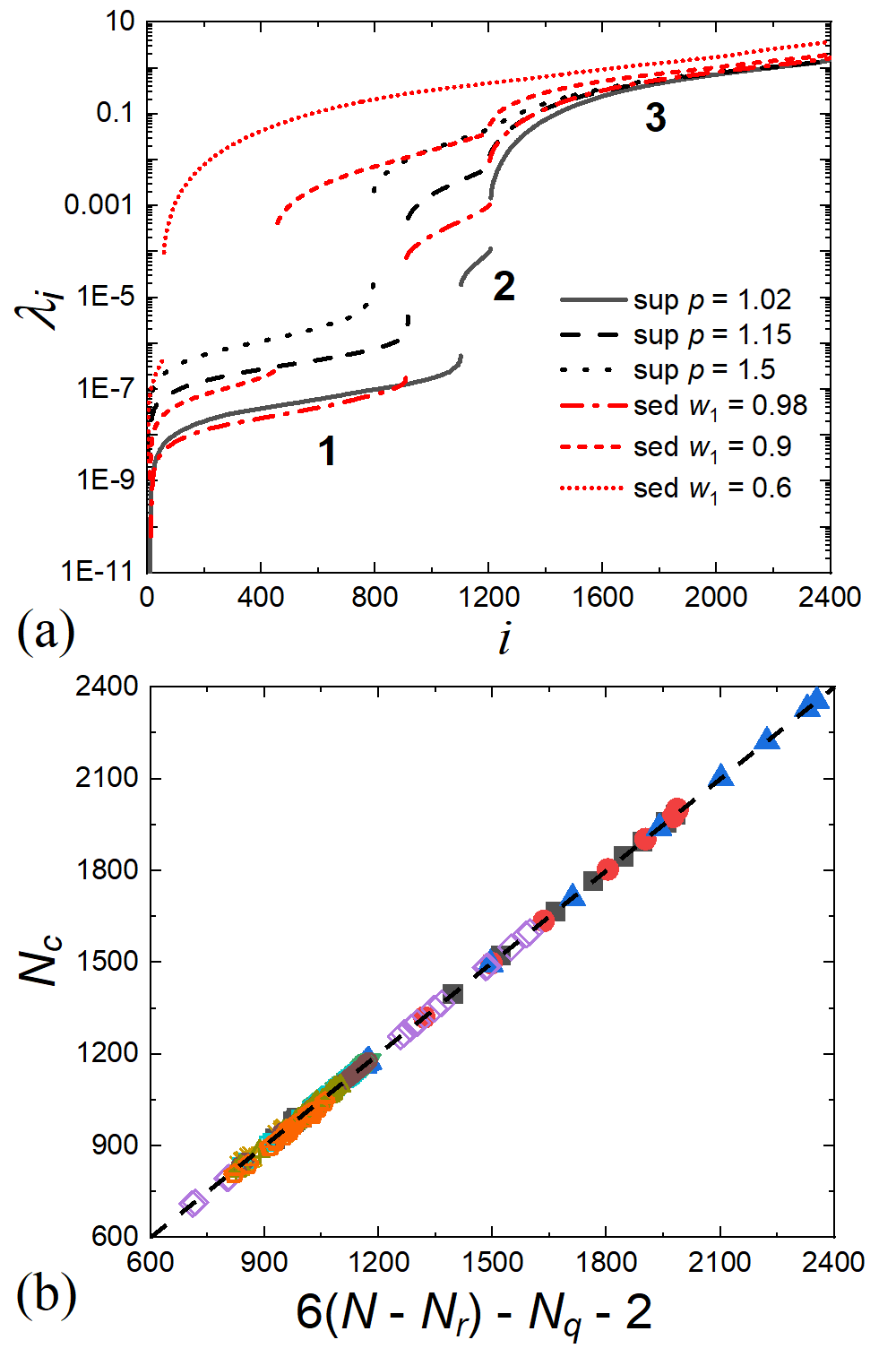}
\caption{(a) Sorted eigenvalues $\lambda_i$ of the dynamical matrix 
for packings of several shapes, including three types of superballs 
($p = 1.02$, $1.15$, and $1.5$) and three types of self-dual ellipsoids 
($w_1 = 0.98$, $0.9$, and $0.6$). Three distinct regimes of 
the spectrum are marked $1$, $2$, and $3$. (b) The number of contacts 
$N_c$ versus $6(N-N_r)-N_q-2$, where $N_r$ is the number of rattler 
particles and $N_q$ is the number of quartic eigenmodes for the packings 
in (a). The dashed 
line has unit slope and passes through the origin.}
\label{spectrum}
\end{figure}

\begin{figure}
\centering
  \includegraphics[height=3.5cm]{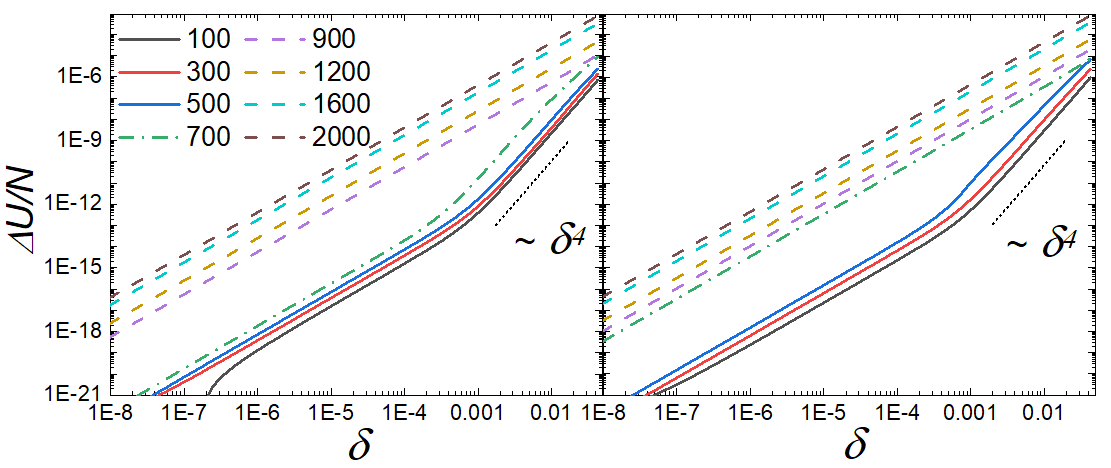}
  \caption{Change in the potential energy per particle $\Delta{U}/N$ 
between the perturbed and unperturbed packing for 
perturbations with amplitdue $\delta$ along several eigenmodes of the 
dynamical matrix for two packings
of superballs with $p = 1.5$ and $z_J = 8.20$ (left) and $9.08$ (right). 
$\Delta U \sim \delta^4$ at large $\delta$ for perturbations along the 
quartic eigenmodes (solid lines), where $\Delta U \sim \delta^2$ for 
perturbations along all other modes (dashed lines).}
\label{quartic}
\end{figure}

In Fig.~\ref{spectrum} (a), we show the eigenvalue spectrum of the
dynamical matrix (Eq.\ref{dm}) sorted from smallest to largest for
packings of $6$ different types of superellipsoids.  As found in
previous studies of packings of ellipsoids, the eigenvalue spectrum
has three distinct regimes.~\cite{schreck2012constraints} For nearly spherical shapes, in regimes
$2$ and $3$, the eigenmodes are purely rotational and translational,
respectively. Regimes $2$ and $3$ merge for systems with sufficiently
large asphericity $\mathcal{A}$. ``Quartic'' modes occur in regime
$1$. When the system is perturbed along an eigenmode in this regime,
the change in the total potential energy $\Delta U$ between the unperturbed 
and pertrubed packings first increases quadratically
with the perturbation amplitude $\delta$, but then scales as
$\delta^4$ beyond a characteristic amplitude $\delta^*$ that scales to
zero with decreasing pressure. (See Fig.~\ref{quartic} (a).) We found
in previous studies of nonspherical particles that the number of
quartic modes $N_q$ matches the deviation in the number of contacts at
jamming onset from the isostatic value, i.e. $N_c = N_c^{\rm iso}
-N_q$, where $N_c^{\rm iso} = d_f (N-N_r) -2$.  We show this result
for packings of superellipsoids generated via protocol $1$ in
Fig.~\ref{spectrum} (b).  This result shows that even though $N_c <
N_c^{\rm iso}$, disordered packings of superellipsoids generated via
protocol $1$ are mechanically stable.

\begin{figure}[t!]
\centering
  \includegraphics[height=10.5cm]{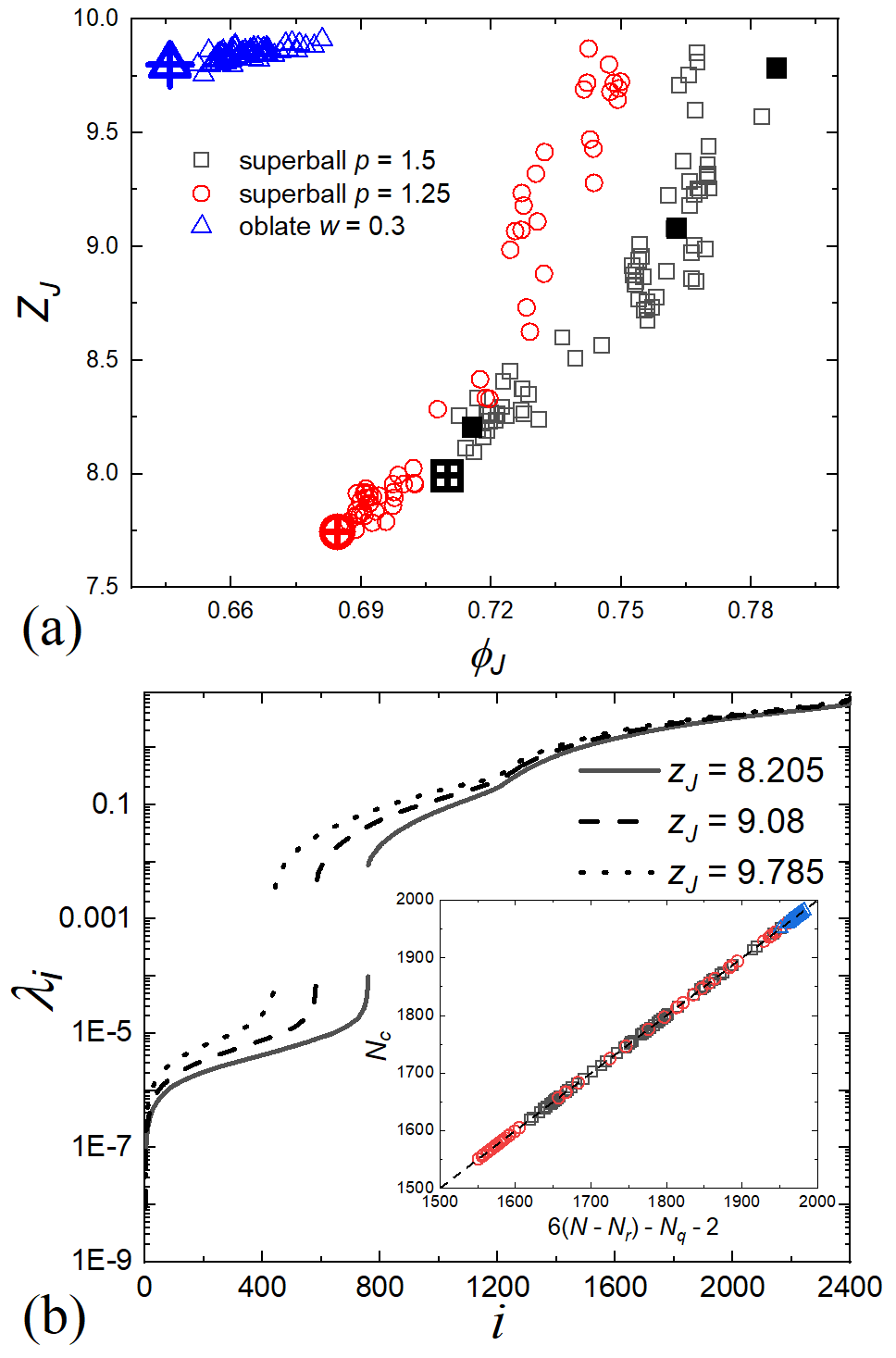}
\caption{(a) The contact number at jamming onset $z_J$ verus the 
packing fraction at jamming onset $\phi_J$ for packings of 
superellipsoidal shapes considered in Fig.~\ref{protocol2} generated 
via protocol $2$.  Results for packings generated using protocol $1$ 
are represented by crosses. (b) Eigenvalues $\lambda_i$ of the dynamical 
matrix sorted 
from smallest to largest for three packings of superballs with 
$p = 1.5$ marked by the solid symbols in (a). The packings possess 
$z_J =8.20$, $9.1$, and $9.8$. The inset shows $N_c$ versus $N_c^{\rm iso}
- N_q$ for all packings of superellipsoids generated via protocol $2$. 
The dashed line has unit slope and passes through the origin.}
\label{z_ordered}
\end{figure}

Is the relationship between the number of contacts and number of
quartic modes the same for packings of nonspherical particles with
significant orientational order?  For example, in ordered systems, it
is possible that some of the $N_c$ contacts are redundant and
therefore do not provide independent constraints to block the degrees of
freedom in the packings.  In Fig.~\ref{z_ordered} (a), we show the
contact number for packings of three types of superellipsoids
generated via protocol $2$ that possess significant global nematic and
cubatic order ({\it c.f.} Fig.~\ref{scatter} (a).)  The contact number
in these systems ($z_J \rightarrow 10$) is much larger than that for
packings generated using protocol $1$.

In Fig.~\ref{z_ordered} (b), we show the eigenvalue spectrum of the
dynamical matrix for three packings of superballs with $p = 1.5$
generated using protocol $2$. As shown previously, the spectrum
includes three regimes with a regime of quartic eigenmodes at the
lowest values.  Further, the crossover in behavior from $\Delta U \sim
\delta^2$ to $\sim \delta^4$ occurs at a similar value of $\delta^*$
that scales to zero with decreasing pressure. (See Fig.~\ref{quartic}
(b).) In the inset of Fig.~\ref{z_ordered} (b), we show the number of
contacts $N_c$ versus the number of quartic modes $N_q$ for all of the
packings generated using protocol $2$.  We find that even with
significant orientational order, the number of quartic modes matches
the deviation in the number of contacts from the isostatic value.
Thus, we find that there are no redundant contacts for hypostatic
packings of superellipsoids with $z_J < z_{\rm iso}$, and $z_J$ determines 
their mechanical stability. 

\begin{figure}
\centering
  \includegraphics[height=6.7cm]{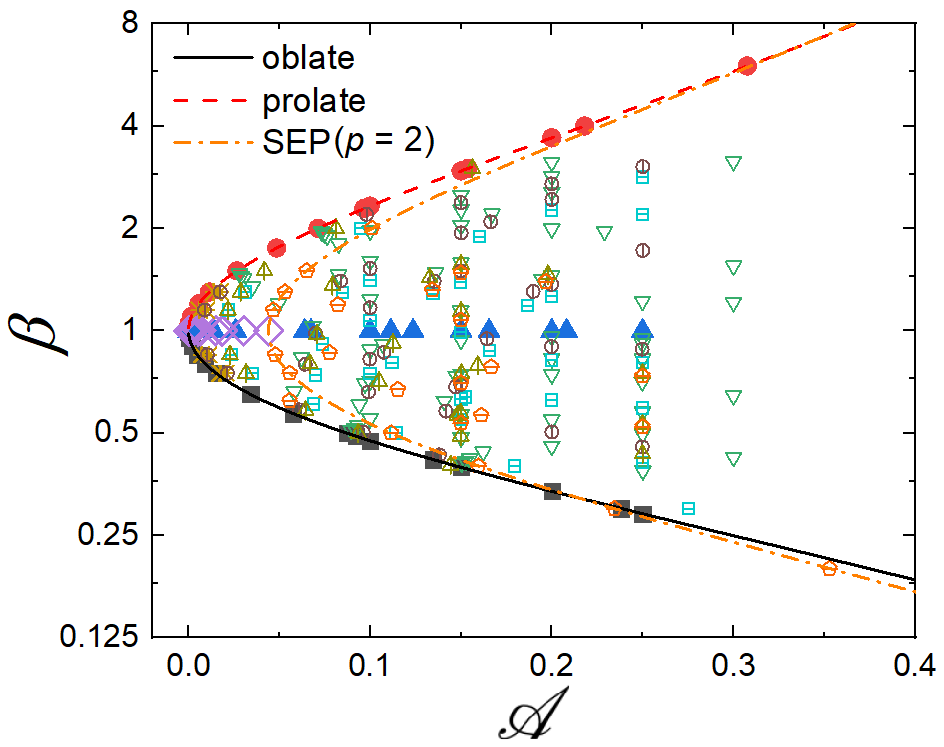}
\caption{The reduced aspect ratio $\beta$ versus the asphericity 
${\cal A}$ for all of the particle shapes studied. The solid and dashed lines 
correspond to Eqs.~\ref{a_oblates} and~\ref{a_prolates} for oblate 
and prolate ellipsoids, respectively.}
\label{beta}
\end{figure} 

\section{Conclusions and Future Directions}
\label{conclusions}

In this article, we carried out computational studies of jammed
packings of frictionless superellipsoids for more than $200$ different
particles shapes in three spatial dimensions. We implemented two
protocols to generate static packings: protocol $1$, which uses
athermal quasistatic compression, and protocol $2$, which includes
thermal fluctuations and compression. Protocol $1$ typically generates
packings with small values of the global nematic and cubatic
orientational order parameters, lower packing fraction $\phi_j$ and
contact number $z_J$ at jamming onset.  In contrast, protocol $2$
allows us to tune the orientational order (as well as $\phi_J$ and
$z_J$) in packings of superellipsoids over a much wider
range compared to those in protocol $1$.

We found several important results. Prior studies of disordered jammed
packings of 2D nonspherical particles have found that the packing
fraction at jamming onset $\phi_J$ for a wide variety of shapes can be
collapsed onto a masterlike curve with respect to a single shape
parameter---the asphericity.~\cite{vanderwerf2018hypostatic} For disordered packings of
superellipsoids in 3D, we find that two shape parameters, e.g. the
asphericity ${\cal A}$ and reduced aspect ratio $\beta$, are required
to collapse $\phi_J$. Additionally, prior studies have found that
packings of nonspherical particles are hypostatic with $z_J < z_{\rm
  iso}$, and the number of missing contacts below the isostatic value
matches the number of quartic eigenmodes of the dynamical matrix.~\cite{schreck2012constraints, vanderwerf2018hypostatic, mailman2009jamming} Most
of these prior studies have considered disordered packings of
nonspherical particles with small values for global measures of
orientational order.  We find that for packings of superellipsoids
with significant orientational order generated via Protocol 2, the
number of missing contacts matches the number of quartic
eigenmodes. Thus, ordered packings of superellipsoids do not possess
any geometrically redundant contacts, and thus the contact number 
$z_J < z_{\rm iso}$ directly determines their mechanical stability.

Our work opens up several new avenues of future research.  First, in
this work, we were able to generate packings of superellipsoids with
tunable orientational order, $\phi_J$, and $z_J$. However, we only
considered packings with $z_J < z_{\rm iso}$. It will be interesting
to generate packings of nonspherical particles with even more order,
where $z_J > z_{\rm iso}$. In this case, do quartic modes still occur
and if so, what determines their number?  Another future research
direction involves packings of {\it frictional} non-spherical
particles.~\cite{neudecker2013jammed, schaller2015local, salerno2018effect, trulsson2018rheology} Packings of frictional spherical particles can occur with
contact numbers that satisfy $d_f +1 < z_J < 2d_f$, where, $d_f = 3$
for spherical particles.~\cite{song2008phase, silbert2010jamming} Prior studies have shown that packings of frictional
nonspherical particles can possess $z_J < d_f + 1$,~\cite{salerno2018effect} where for example $d_f
= 5$ for axisymmetric particles. Do these packings possess quartic modes, 
and if so, how many?  It is clear that much more work is needed to understand the number of contacts that are required to determine the mechanical stability 
of packings of frictional, nonspherical particles.

\section*{Conflicts of interest}
There are no conflicts to declare.

\section*{Appendix A: Variation of the shape parameters $\beta$ and ${\cal A}$}
\label{A}

In this Appendix, we show the range of reduced aspect ratio $\beta$
and asphericity ${\cal A}$ that can be achieved for superellipsoidal
particle shapes. For oblate and prolate ellipsoids, the asphericity
${\cal A}(\beta)$ can be written explcity. For oblate ellipsoids, we
find
\begin{equation}
{\cal A}(\beta) = \frac{2\beta^{2/3}}{1+\frac{\beta^2}{\sin\gamma}\ln \left(\frac{1+\sin\gamma}{\cos\gamma} \right)},
\label{a_oblates}
\end{equation}
where $\gamma = \cos^{-1} \beta$. For prolate ellipsoids, ${\cal A}(\beta)$ 
can be expressed as 
\begin{equation}
{\cal A}(\beta) = \frac{2\beta^{2/3}}{ 1+\frac{\beta \alpha}{\sin\alpha} },
\label{a_prolates}
\end{equation}
where $\alpha = \cos^{-1} \beta^{-1}$. In Fig.~\ref{beta}, we plot the 
relations between $\beta$ and ${\cal A}$ for oblate and prolate ellipsoids, 
as well as $\beta({\cal A})$ for superellipsoids with $p=2$.  We find 
that these curves serve as upper and lower bounds for the shape 
parameters of all other shapes that we study for ${\cal A} < 0.35$.   

\section*{Appendix B: Local nematic and cubatic order parameters}
\label{B}

In the main text, for example in Figs.~\ref{scatter} (a)
and~\ref{protocol2} (a), we showed results for the global nematic
$S_2$ and cubatic $C_4$ order parameters for packings of
superellipsoids. In these figures, we also show example packings from
the simulations with the particles colored according to the value of
the local nematic and cubatic order parameters. The local nematic
order parameter $S_2^{\rm loc}$ is defined analogously to Eq.~\ref{s2}
as the largest eigenvalue of the $3 \times 3$ matrix:
\begin{equation}
\label{s2_local}
S^{\rm loc}_{\alpha \beta} = \frac{3}{2} \left\langle {\hat s}_{\alpha i} \cdot
{\hat s}_{\beta j} \right\rangle_j
- \frac{\delta_{\alpha \beta}}{2},
\end{equation}
where $\langle .\rangle_j$ averages over particles $j$ that overlap 
particle $i$. 

\begin{figure}
\centering
  \includegraphics[height=3.5cm]{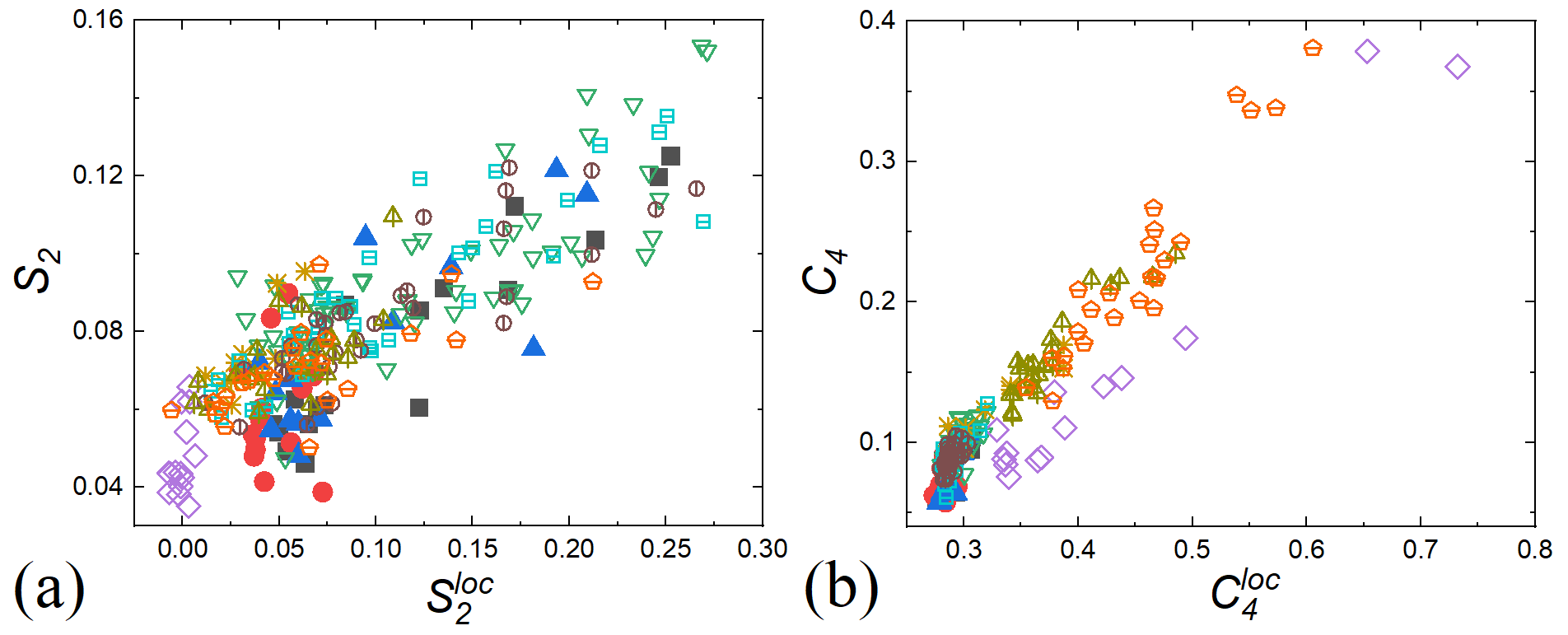}
  \caption{(a) The global nematic order parameter $S_2$ plotted 
versus the local nematic order parameter $S_2^{\rm local}$ for all 
particle shapes 
considered. (b) The global cubatic order parameter $C_4$ plotted 
versus the local cubatic order parameter $C_4^{\rm loc}$ for all 
particle shapes considered.}
\label{global_vs_local}
\end{figure}

To define the local cubatic order parameter $C_4^{\rm loc}$ for
particle $i$, we first calculate
\begin{equation}
P_4({\hat u}_i, {\hat u}_j) =\frac{1}{8}\left( 35 [{\hat u}_j \cdot {\hat{\emph{\textbf{u}}}}_i]^4 - 30 [ {\hat u}_j \cdot {\hat{\emph{\textbf{u}}}}_i]^2 + 3 \right),
\label{cubatic_local}
\end{equation}
where ${\hat u}_j$ is a unit vector aligned with one of the three
orientations of the semi-major axes for particle $j$ that overlaps
particle $i$. We first select the ${\hat u}_i$ orientation along one
of the three semi-major axes that maximizes $P_4({\cal u}_i, {\cal
  u}_j)$ for a given ${\cal u}_j$. We then average $P_4^{\rm
  max}({\hat u}_j)$ over all particles $j$ that overlap $i$. The local
cubatic order parameter $C_4^{\rm loc}$ is defined as the maximum over
the three orientations for ${\cal u}_j$. We plot the global versus the
local orientational order parameters in
Fig.~\ref{global_vs_local}. For the nematic and cubatic order, the
global and local order grow proprotionately.
  
\section*{Appendix C: Gaussian curvature at contact points}
\label{C}

\begin{figure}
\centering
  \includegraphics[height=6.7cm]{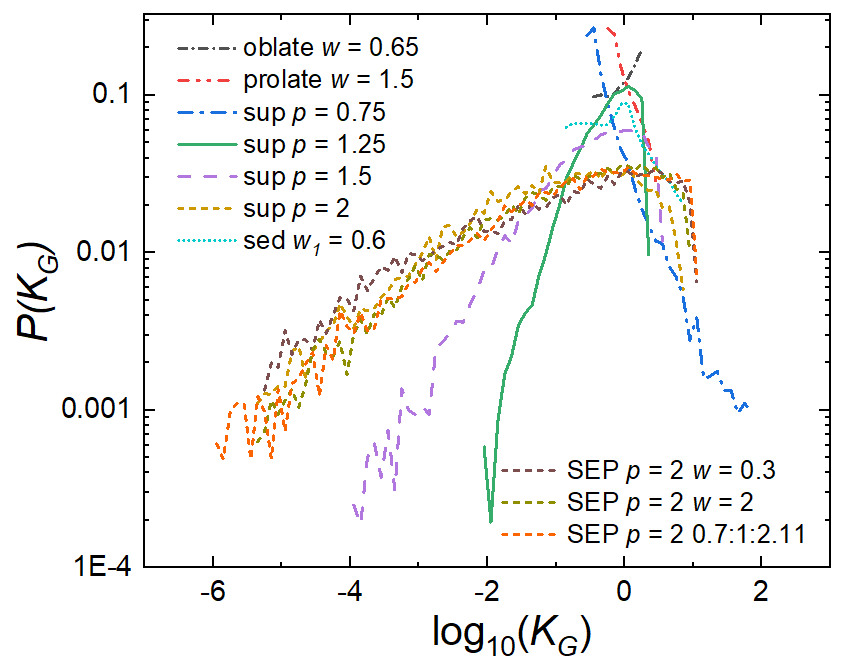}
  \caption{Probability distribution of the scaled Gaussian curvature 
at each interparticle contact $P(K_G)$ in superellipsoid packings 
generated via protocol $1$.}
\label{curvature}
\end{figure}

In this Appendix, we show that for packings of superellipsoids with
small contact numbers at jamming onset, the Gaussian curvature
${\overline K}_G$ at the points of contact are typically small,
suggesting that two flat contacting surfaces can constrain multiple
rotational degrees of freedom.  In Fig.~\ref{curvature}, we show the
probability distribution $P(K_G)$, where $K_G = {\overline K}_G
(abc)^{2/3}$, for superellipsoid packings generated via protocol
$1$. (Note that each contact point contributes two $K_G$ values.) We
find that $P(K_G)$ for packings of cube-like superellipsoids,
e.g. with $p = 2$ and small $z_J$, possess a wide tail that extends to
small values of $K_G$.  For other particle shapes, such as oblate and
prolate ellipsoids, $P(K_G)$ is much narrower and does not extend to
small values of $K_G$.

\section*{Acknowledgements}

We acknowledge support from NSF Grants Nos. CBET-1605178 (K. V. and
C. O.) and CMMI-1463455 (M. S.) and the China Scholarship Council Grant
No.  201806010289 (Y. Y.). This work was also supported by the High
Performance Computing facilities operated by Yale's Center for
Research Computing. In addition, we acknowledge the High-performance Computing
Platform of Peking University. We also thank S. Li at Peking University
for helpful discussions.  



\balance

\renewcommand\refname{References}



\bibliography{supelp_softmat.bib}
\bibliographystyle{rsc} 
\end{document}